# Deep-Subwavelength and Broadband Quarter-Wave Retardation in Ultrathin Hyperbolic MoOCl$_2$

Georgy Ermolaev[1#], Adilet Toksumakov[1#], Valeria Maslova[1,2], Aleksandr Slavich[1], Anton Minnekhanov[1], Gleb Tselikov[1], Nikolay Pak[1], Andrey Vyshnevyy[1], Aljoscha Söll[3], Zdeněk Sofer[3], Aleksey Arsenin[1], Kostya S. Novoselov[4,5,6*], and Valentyn Volkov[1*]

[1]Emerging Technologies Research Center, XPANCEO, Internet City, Emmay Tower, Dubai, United Arab Emirates

[2]C.H.O.S.E. (Center for Hybrid and Organic Solar Energy), Electronic Engineering Department, University of Rome Tor Vergata, Via del Politecnico 1, 00118, Rome, Italy

[3]Department of Inorganic Chemistry, University of Chemistry and Technology Prague, Prague 6, Czech Republic

[4]National Graphene Institute (NGI), University of Manchester, Manchester M13 9PL, UK

[5]Department of Materials Science and Engineering, National University of Singapore, Singapore 03-09 EA, Singapore

[6]Institute for Functional Intelligent Materials, National University of Singapore, Singapore 117544, Singapore

[#]These authors contributed equally to this work

*Correspondence should be addressed to the e-mails: vsv@xpanceo.com and kostya@nus.edu.sg

## Abstract

The miniaturization of polarization-controlling optical components is one of the central pursuits in nanophotonics. While traditional anisotropic materials require large propagation lengths to achieve the desired phase shifts, metasurfaces mitigate this size constraint but often introduce narrow operational bandwidths and high fabrication complexities. To bridge this gap, we introduce MoOCl$_2$ as a promising material for ultracompact and broadband phase retardation. Building on its giant optical anisotropy, we experimentally demonstrate MoOCl$_2$ quarter-wave plates with thicknesses of 77 nm and 98 nm. These flakes exhibit achromatic quarter-wave retardation across broad visible (445 – 525 nm) and near-infrared (730 – 945 nm) spectral windows, surpassing the fundamental thickness and bandwidth limitations of both conventional optical materials and artificial nanostructures. Moreover, MoOCl$_2$ waveplates demonstrate up to $\lambda/4500$ retardance tolerance at central wavelengths. As a result, this study establishes MoOCl$_2$ as a building block for ultracompact polarization optics.

**Keywords:** van der Waals materials, hyperbolic materials, MoOCl$_2$, anisotropy, waveplates

Precise control over the polarization state of light is fundamental to modern photonics, with a wide array of applications ranging from high-resolution biosensing[1,2] to augmented reality[3,4], optical communication[5,6], and quantum information processing[7,8]. At the heart of these technologies lies the concept of optical phase retardation in waveplates, which manipulate light polarization by introducing a controlled phase shift between orthogonal electric field components[9,10]. Traditionally, this optical phase retardation is achieved using naturally birefringent crystals, such as calcite[11], quartz[12], and rutile[13], which exhibit different refractive indices along their principal optical axes. However, the relatively weak optical anisotropy inherent to these conventional bulk materials necessitates propagation lengths on the order of tens to hundreds of micrometers to accumulate a sufficient phase difference for a quarter-wave or half-wave shift[12,14]. This optically thick requirement poses a major barrier to the ongoing quest to miniaturize next-generation nanophotonics devices[6,15,16].

To overcome the size limitations of traditional crystals, substantial efforts have been directed toward artificial subwavelength structures, including form-birefringent metamaterials[17,18] and plasmonic[19,20] or dielectric[21,22] metasurfaces. While these engineered structures have successfully reduced the footprint of polarization-controlling devices, they often suffer from significant drawbacks, such as complex and costly nanoscale fabrication processes, narrow operational bandwidths, and considerable scattering optical losses, especially in the visible spectral range[17–22]. Consequently, there is an urgent need to identify natural materials that intrinsically possess giant optical anisotropy, enabling deeply subwavelength phase retardation without the need for nanolithography.

Recent advances in materials science have identified van der Waals (vdW) crystals as a promising platform for overcoming these technological limitations[23–31]. Characterized by strong in-plane covalent bonding and weak vdW forces, these low-symmetry materials naturally host high values of optical anisotropy[32]. This natural anisotropy has facilitated the realization of subwavelength quarter-wave plates based on $As_2S_3$[10], $\alpha$-$MoO_3$[33], $CrSBr$[34], $NbOCl_2$[35], and low-dimensional red phosphorous[36] and ferrocene[37] crystals. Still, achieving broadband achromatic and deeply subwavelength phase retardation using these materials remains a challenge.

In this context, layered molybdenum oxydichloride ($MoOCl_2$) is a highly promising material system[5,38–48]. Crystallizing in a low-symmetry monoclinic lattice, $MoOCl_2$ is characterized by parallel Mo-O chains separated by halogen atoms[49], which imparts a quasi-one-dimensional (quasi-1D) nature to its structural and electronic framework[38]. Recent investigations of $MoOCl_2$ have unveiled its record broadband birefringence[38] and the strongly correlated nature of its electrons[50], showcasing phenomena such as Fermi-liquid behaviour[51], colossal magnetoresistance[51], and highly directional plasmon polaritons[43,46]. From an optical perspective, this profound structural asymmetry manifests as a broadband in-plane hyperbolic response extending from the visible to the infrared wavelengths[38]: $MoOCl_2$ crystal exhibits metallic behavior ($\varepsilon_{1a} < 0$) along the crystallographic *a*-axis and dielectric behavior ($\varepsilon_{1b}, \varepsilon_{1c} > 0$) along the orthogonal crystallographic *b*-axis and *c*-axis. Although these highly anisotropic properties have been widely recognized[5,38–48], the potential of $MoOCl_2$ to precisely modulate light phase, specifically its ability to function as an ultracompact broadband waveplate, has not been fully explored.

Here, we show that $MoOCl_2$ provides a natural platform for deep subwavelength ($\lambda/d > 10$) and broadband (445 – 525 nm and 730 – 945 nm) achromatic quarter-wave retardation. Our findings establish that the total phase retardation in ultrathin $MoOCl_2$ flakes is not solely dictated by the classical phase accumulation, but is significantly modified by Fabry-Pérot interference effects within $MoOCl_2$ nanoscale cavity. Using this model, we theoretically and experimentally demonstrated deep subwavelength $MoOCl_2$ waveplates with thicknesses of just 77 nm and 98 nm, capable of efficiently converting linearly polarized light into a circularly polarized state. These characteristics of MoOCl$_2$ outperform both classical and vdW anisotropic materials and artificial metasurface-based waveplates. Therefore, our results establish MoOCl$_2$ not only as a model hyperbolic vdW material, but also as a practical building block for ultracompact broadband polarization optics.

To evaluate the potential of MoOCl$_2$ for ultracompact polarization elements, we first recall its fundamental crystallographic structure with its macroscopic optical response (Figure 1). MoOCl$_2$ is a layered vdW material that crystallizes in a low-symmetry monoclinic lattice, shown in Figure 1a. The intralayer atomic arrangement is characterized by strongly bonded quasi-1D Mo-O-Mo chains extending along the crystallographic *a*-axis, which are laterally separated by weakly interacting chlorine atoms along the crystallographic *b*-axis. This profound structural asymmetry translates directly into a highly anisotropic biaxial dielectric tensor. As a result, the material's dielectric function ellipsoid undergoes a drastic transformation with the wavelength change (Figure 1b). Most notably, the high-density Mo-O chains promote a metallic optical response ($\varepsilon_{1a} < 0$) along the crystallographic *a*-axis, while the orthogonal *b*- and *c*-axes retain a dielectric character ($\varepsilon_{1b,c} > 0$).

The optical manifestation of this metallic-dielectric duality is captured in the dispersion of the optical constants from the ultraviolet to the near-infrared spectra (Figure 1c). The refractive index (*n*) and extinction coefficient (*k*) spectra reveal a broad Drude-like metallic tail along the crystallographic *a*-axis, contrasting sharply with the low-loss dielectric behavior along the crystallographic *b*- and *c*-axes. This extreme disparity yields a giant in-plane birefringence ($\Delta n = n_a - n_b$), as well as a large linear dichroism ($\Delta k = k_a - k_b$), seen in Figure 1d. Such a great difference in the phase velocity of orthogonal electric field components provides an intrinsic mechanism for accumulating significant optical phase retardation over extremely short propagation distances.

The technological advantage of this giant birefringence becomes apparent when estimating the minimum material thickness required to achieve a quarter-wave phase shift $\delta$ ($\delta = \pi/2 = 90°$). In the classical approximation, the thickness $d$ of a quarter-wave plate is inversely proportional to the material's birefringence via the formula $d = \lambda/(4\Delta n)$. Figure 1e benchmarks the required quarter-wave thickness of MoOCl$_2$ against a comprehensive library of state-of-the-art birefringent materials. This includes traditional bulk crystals like rutile and calcite, as well as emerging highly anisotropic vdW materials, such as As$_2$S$_3$, CrSBr, and NbOCl$_2$. Clearly, with conventional materials, polarization optics are constrained to the optically thick regime ($d > \lambda$), while the giant birefringence of MoOCl$_2$ dramatically compresses the required optical path. Strikingly, MoOCl$_2$ enables quarter-wave retardation at physical thicknesses that are strictly subwavelength ($d < \lambda$) across a broad spectral range (Figure 1e).

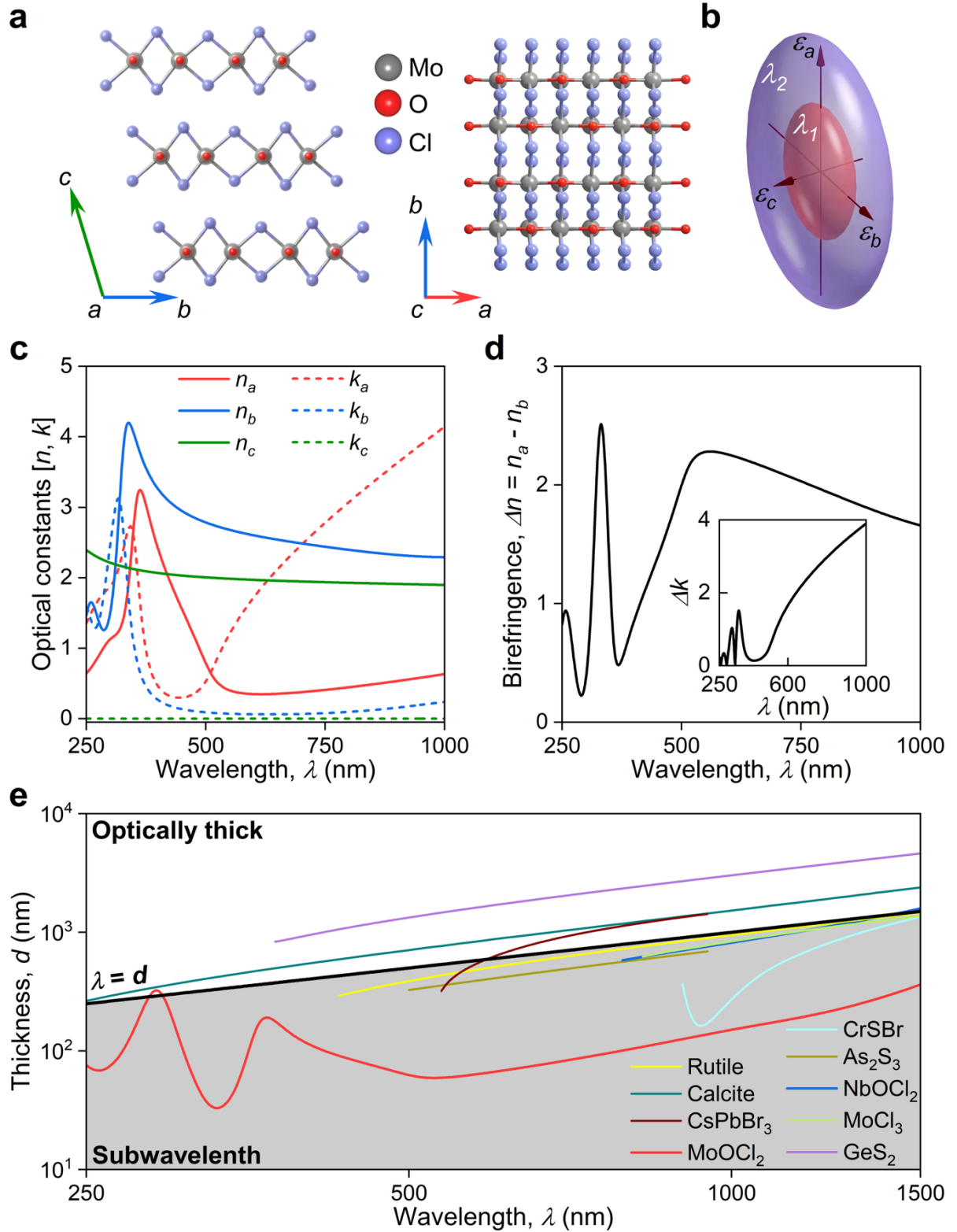

**Figure 1. Crystal structure and giant optical anisotropy of MoOCl$_2$. a,** Layered MoOCl$_2$ crystal structure viewed along the *a*-axis (left panel) and the *c*-axis (right panel). Grey, red, and violet spheres denote Mo, O, and Cl atoms, respectively. **b,** Illustration of the anisotropic dielectric function ellipsoid at two distinct wavelengths ($\lambda_1$ and $\lambda_2$), highlighting MoOCl$_2$ triaxial nature of the complex dielectric tensor. **c,** Dispersion of the optical constants, including refractive index (*n*, solid lines) and extinction coefficient (*k*, dashed lines), along the *a*, *b*, and *c*-axes from the ultraviolet to the near-infrared region. **d,** In-plane birefringence ($\Delta n = |n_a - n_b|$) as a function of wavelength, demonstrating giant optical anisotropy. The inset shows the corresponding linear dichroism ($\Delta k = |k_a - k_b|$). **e,** Comparison of the required material thickness ($d = \lambda/(4\Delta n)$ in classical approximation) for quarter-wave retardation across various birefringent materials. The solid black line marks the boundary, where the thickness equals the operating wavelength ($d = \lambda$). Unlike traditional materials that operate in the optically thick regime, MoOCl$_2$ achieves exceptional phase retardation at subwavelength thicknesses across a broad spectral range.

To harness the extraordinary optical anisotropy of MoOCl$_2$ for ultracompact polarization manipulation, we theoretically calculate the design and performance of a subwavelength quarter-wave plate based on MoOCl$_2$. The proposed device architecture, conceptually illustrated in Figure 2a, comprises an ultrathin MoOCl$_2$ flake placed on a glass substrate. In such highly anisotropic subwavelength structures, the accumulated optical phase retardance ($\delta$) cannot be accurately captured by the classical propagation model ($\delta = (2\pi/\lambda)d\Delta n$) alone. Instead, the total phase difference is governed by a combination of this conventional phase accumulation term, which scales with the intrinsic birefringence ($\Delta n$) and thickness ($d$), and a substantial contribution from Fabry-Pérot interference effects within the nanoscale cavity of MoOCl$_2$ flake itself.

To rigorously quantify the optical response of these waveplates, we computed the transmittance spectra for incident light polarized along the crystallographic *a*- and *b*-axes of MoOCl$_2$ for flake thicknesses ranging from 50 to 100 nm (Figure 2b). Clearly, the transmittance for the two orthogonal electric field components differs significantly because of the profound linear dichroism of MoOCl$_2$. Consequently, generating a circular polarization state at the output requires the incident linearly polarized light to be oriented at a polarization angle ($\varphi$) that compensates for this amplitude disparity. As depicted in Figure 2c, this optimal incident polarization angle $\varphi$ exhibits a distinct spectral dependence and diverges noticeably from the conventional 45-degree alignment utilized in standard waveplates.

The synergy between the material's giant natural birefringence and the internal Fabry-Pérot resonances drastically reduces the physical footprint required to achieve a 90-degree phase shift. In Figure 2d, we plot the minimal MoOCl$_2$ thickness necessary for quarter-wave operation across the visible spectrum. The exact analytical solution based on the transfer matrix calculations (see Methods section in Supplementary Information) demonstrates a noticeable improvement in miniaturization compared to the classical approximation, underscoring the critical role of cavity-enhanced phase shifts in ultrathin media. This exceptional degree of miniaturization is further highlighted by the phase retardation efficiency, defined as the ratio of the operational vacuum wavelength ($\lambda$) to the minimal required device thickness ($d$) in Figure 2e. The device maintains remarkably high efficiency metrics, penetrating even beyond the deep subwavelength ($\lambda/d > 10$) regime. Finally, the calculated phase retardance spectra for selected subwavelength thicknesses ($d = 50 - 100$ nm) confirm that the required quarter-wave phase shift ($\delta = 90°$) can be precisely targeted at specific wavelengths across the visible range by simply fine-tuning the flake thickness.

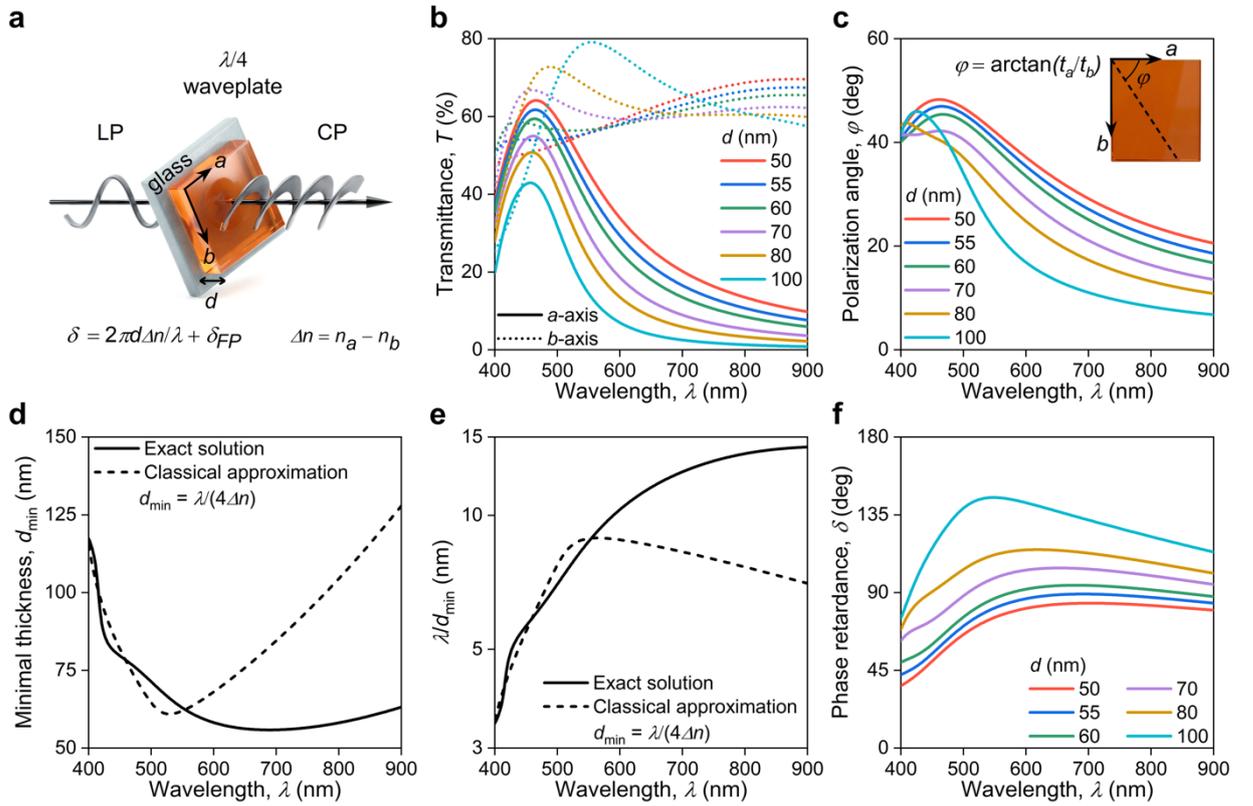

**Figure 2. Theoretical design and performance of subwavelength MoOCl₂ waveplates. a,** Schematic illustration of MoOCl₂-based subwavelength quarter-wave plate ($\delta = \pi/2$) on a glass substrate, converting linearly polarized (LP) light into circularly polarized (CP) light. The total phase retardance ($\delta$) is defined by both the classical phase accumulation term ($\delta_{\text{classic}} = 2\pi d\Delta n/\lambda$) and the Fabry-Pérot interference contribution ($\delta_{\text{FP}}$). **b,** Calculated transmittance ($T$) spectra for the incident light polarized along the crystallographic a-axis (solid lines) and b-axis (dotted lines) of MoOCl₂ flakes with thicknesses ($d$) ranging from 50 to 100 nm. **c,** Spectral dependence of the incident light polarization angle ($\varphi$), required for equal electric field amplitudes of the transmitted light in order to obtain the circular polarization. **d,** Minimal MoOCl₂ thickness ($d_{\min}$) required to achieve quarter-wave operation. The exact analytical solution (solid line) demonstrates a significant improvement compared to the classical approximation ($t_{\min} = \lambda/(4\Delta n)$, dashed lines). **e,** MoOCl₂ phase retardation efficiency, expressed as the ratio of the operating wavelength to the minimal required thickness ($\lambda/t$). **f,** Calculated phase retardance spectra for the selected subwavelength MoOCl₂ thicknesses.

To experimentally validate our theoretical concept, we used a micro-polarimetry setup (see Methods section in Supplementary Information) to quantify the optical phase retardation of ultrathin MoOCl₂ flakes. As schematically illustrated in Figure 3a, linearly polarized light falls onto MoOCl₂ flake, and the transmitted light is subsequently collected and passed through an analyzer to determine its final polarization state. For a comprehensive analysis, we focused on two representative MoOCl₂ flakes with the measured thicknesses of $d = 77$ nm (Figure 3b) and $d = 98$ nm (Figure 3c).

Using the setup in Figure 3a, we recorded broadband polarized transmittance as a function of both the incident wavelength $\lambda$ and the polarization angle $\theta_P$ (the analyzer angle $\theta_A$ was set to the parallel configuration, $\theta_P = \theta_A = \theta$) for 77 nm (Figure 3d) and 98 nm (Figure 3e) thick flakes. From these spectra, we extract the experimental phase retardance $\delta(\lambda)$ wavelength dependence by leveraging the formula[31] for the polarized transmittance $T(\theta, \lambda) = a^2 \cos^4(\theta - \varphi) + b^2 \sin^4(\theta - \varphi) + 2ab \cos^2(\theta - \varphi) \sin^2(\theta - \varphi) \cos(\delta)$, where $a^2$ and $b^2$ are transmittance along the crystallographic a- and b-axes, respectively, and $\varphi$ is the polarization angle corresponding to the crystallographic a-axis ($\varphi + 90°$ corresponds to the crystallographic b-axis). The theoretical and experimental curves of $\delta(\lambda)$ are presented in Figures 3f-g, which show a good agreement with each other. Moreover, the behavior of $\delta(\lambda)$ clearly deviates from the classical approximation ($\delta = (2\pi/\lambda)d\Delta n$), confirming our earlier assertion that Fabry-Pérot cavity resonances enhance the overall phase shift. From Figure 3g, we see that for 77 nm-thick MoOCl$_2$, quarter-wave retardation ($\delta = 90°$) is achieved at dual wavelengths of 482 nm and 818 nm. Similarly, 98 nm-thick MoOCl$_2$ operates as a quarter-wave plate at 420 nm and 787 nm.

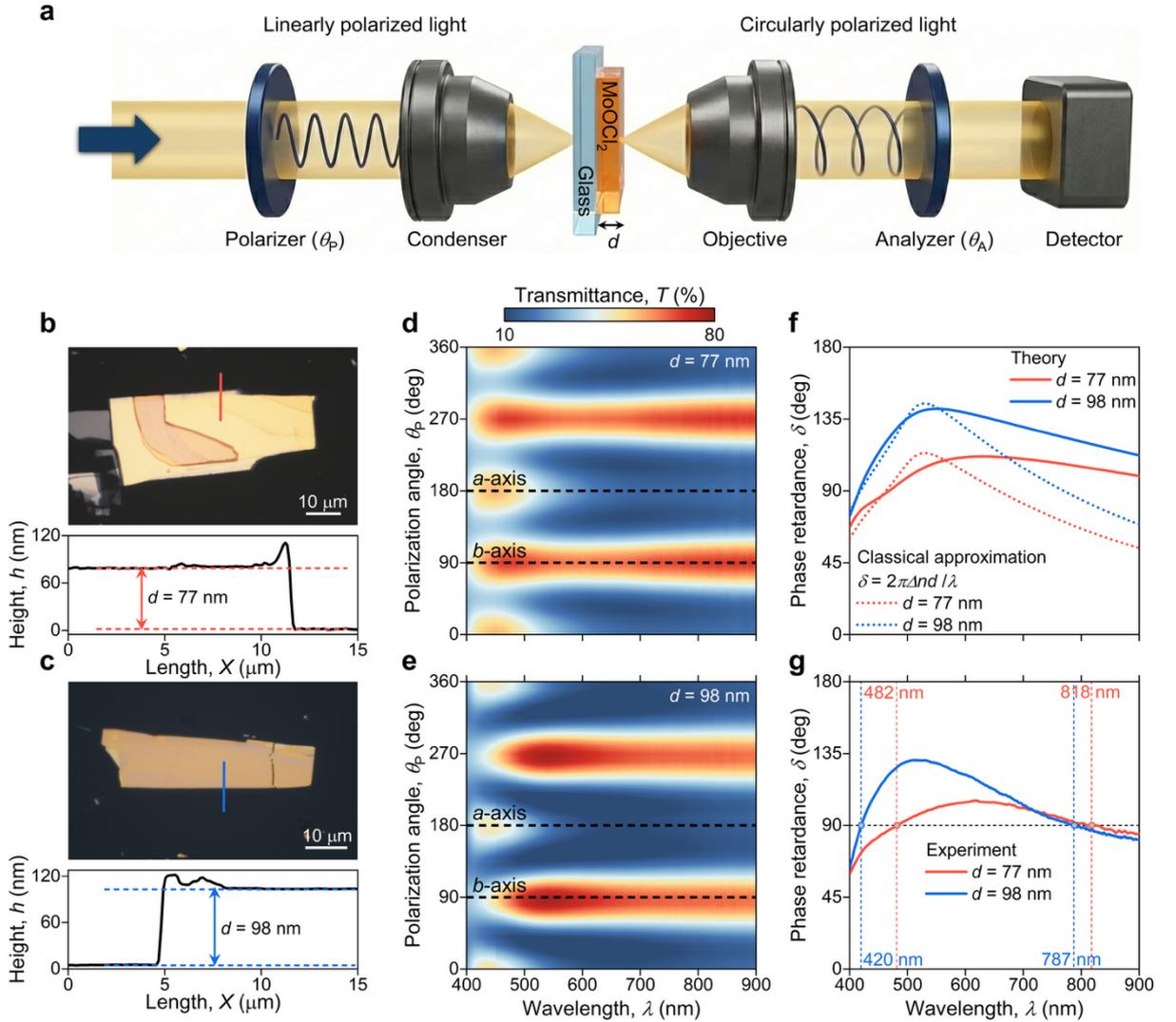

**Figure 3. Experimental realization of subwavelength MoOCl$_2$ waveplates. a,** Schematic of the experimental setup used to measure optical phase retardation of MoOCl$_2$ flake on a glass substrate. **b-c,** Optical microscopy images (top) and corresponding atomic-force microscopy heights profiles (bottom) for MoOCl$_2$ flakes with measured thickness of (**b**) $d$ = 77 nm and (**c**) $d$ = 98 nm. **d-e,** Heatmaps of the measured transmittance as a function of the incident wavelength and polarization angle for (**d**) 77 nm and (**e**) 98 nm-thick flakes. The dashed horizontal lines denote the polarization along the crystallographic a-axis and b-axis. **f,** Calculated phase retardance spectra for the two flake thicknesses. Solid lines represent the exact theoretical model, whereas dotted lines denote the classical approximation ($\delta = 2\pi\Delta nd/\lambda$). **g,** Experimentally determined phase retardance spectra for 77 nm and 98 nm-thick MoOCl$_2$ flakes. Dashed lines highlight the precise wavelengths where exact quarter-wave retardation (90°) is achieved for each respective thickness.

To conclusively verify the generation of circularly polarized light in MoOCl$_2$ quarter-wave plates, we performed their polarimetry analysis. Traditionally, for this verification, we need to shine the linearly polarized light at $\theta_P = 45°$ polarization on waveplates and check that the transmitted light intensity does not depend on the analyzer angle $\theta_A$. However, in MoOCl$_2$ case, we need to find the analyzer angle that compensates for MoOCl$_2$ dichroism. By independently rotating the polarizer $\theta_P$ and analyzer $\theta_A$ angles, we mapped the normalized transmittance to identify this $\theta_P$ corresponding to circular polarization generation. For 77 nm-thick MoOCl$_2$ operating at the experimentally determined quarter-wave wavelength of 482 nm, the transmittance $T(\theta_P, \theta_A)$ map reveals that at $\theta_P = 43.2°$ the transmitted intensity remains constant regardless of $\theta_A$ (Figure 4a). This uniform intensity under analyzer rotation serves as the hallmark signature of a circularly polarized state. Similarly, at the near-infrared quarter-wave condition of 818 nm for the same 77 nm flake, this invariant transmittance is achieved when the incident light is polarized at $\theta_P = 29.0°$ (Figure 4b). Furthermore, we observed highly consistent behavior in 98 nm-thick MoOCl$_2$. At quarter-wave operation wavelength of 420 nm, the condition for circular polarization is satisfied with the incident polarization angle $\theta_P = 45.0°$ (Figure 4c). Meanwhile, for 787 nm, the optimal polarization angle $\theta_P = 25.0°$ (Figure 4d). At these optimal polarization angles $\theta_P$, we can calculate the retardance tolerance $(\Delta)^{35}$ of our waveplates using the minimum $I_{min}(\theta_P)$ and maximum $I_{max}(\theta_P)$ intensities (in ideal case of circular polarization, $I_{min}(\theta_P) = I_{max}(\theta_P)$) at $\theta_P$ via formula $\Delta = \lambda |90 - 2\tan^{-1}(\sqrt{I_{min}/I_{max}})|/360$. For 98 nm waveplate, we obtain $\Delta(787\ nm) = \lambda/100$ and $\Delta(420\ nm) = \lambda/200$, which is comparable with commercial waveplates tolerance[35]. Meanwhile, for 77 nm waveplate, we obtain $\Delta(818\ nm) = \lambda/700$ and $\Delta(482\ nm) = \lambda/4500$, which overcomes even recently shown waveplates[35] based on NbOCl$_2$ with retardance tolerance $\lambda/600$ and $\lambda/300$. Altogether, these polarimetry measurements unambiguously confirm the operation of MoOCl$_2$ flakes as quarter-wave plates.

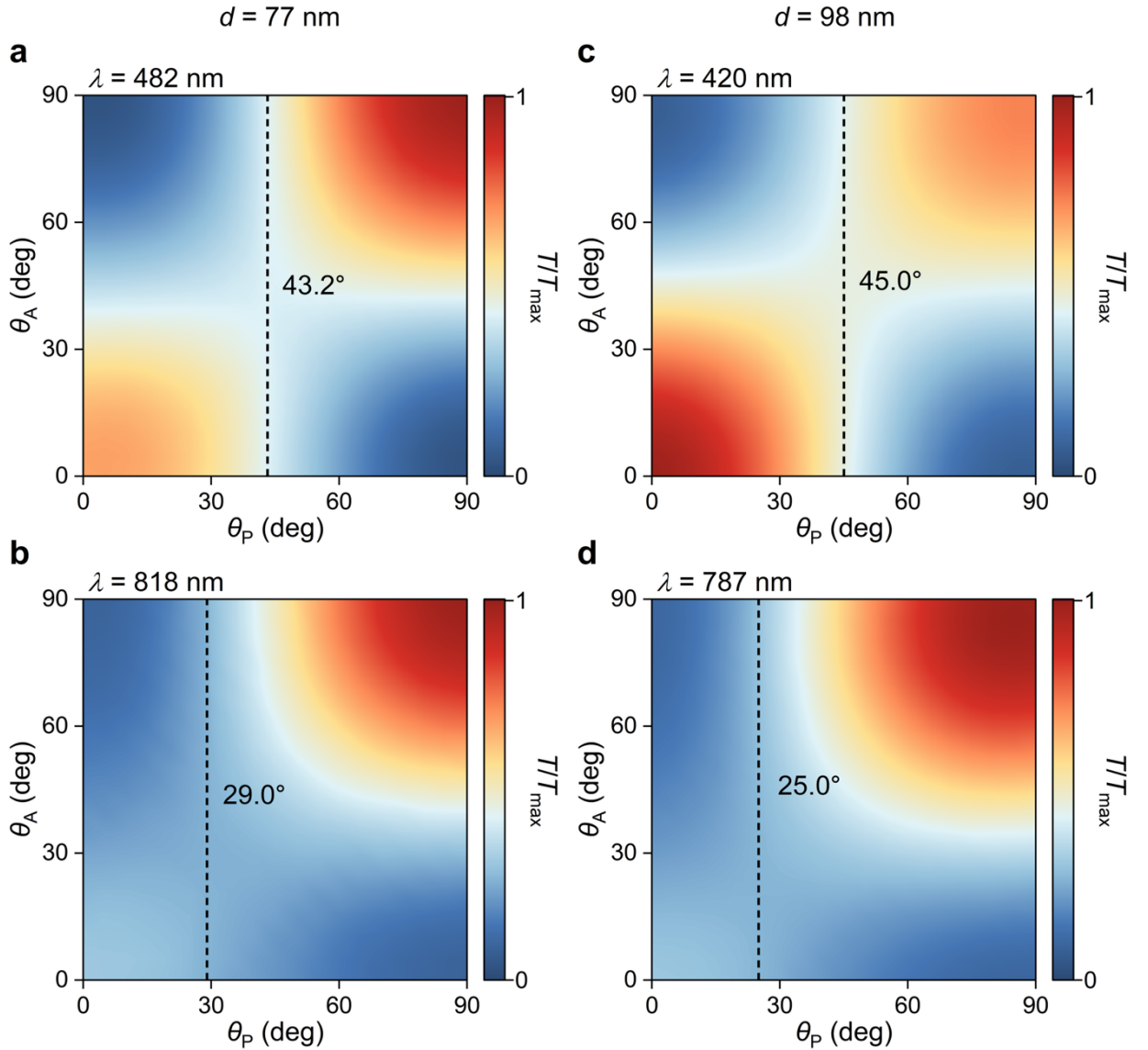

**Figure 4. Polarimetry analysis confirming circular polarization generation in subwavelength MoOCl$_2$. a-b,** Normalized transmittance heatmaps as a function of the polarizer and analyzer angles for 77 nm-thick MoOCl$_2$ flake. Measurements are taken at the experimentally determined quarter-wave operation wavelengths of (a) $\lambda$ = 482 nm and (b) $\lambda$ = 818 nm. **c-d,** Corresponding normalized transmittance maps for 98 nm-thick MoOCl$_2$ flake at its respective quarter-wave operation wavelengths of (**c**) $\lambda$ = 420 nm and (**d**) $\lambda$ = 787 nm. The vertical dashed lines in all panels indicate the specific incident polarization angle ($\theta_P$) required to achieve circularly polarized light. At these specific $\theta_P$ values, the transmitted intensity remains constant regardless of the analyzer angle, which is the hallmark signature of circular polarization.

To contextualize the exceptional phase-modulation capabilities of MoOCl$_2$, we benchmark the minimum material thickness required to achieve quarter-wave retardation against a diverse library of the state-of-the-art anisotropic materials[10,33–37] and nanostructures[21,22] (Figure 5a). From Figure 5a, we see that MoOCl$_2$ provides the smallest thickness for a quarter-wave operation. This level of miniaturization is further quantified by examining the wavelength-to-thickness ratio ($\lambda/d$), where MoOCl$_2$ overcomes even a deep subwavelength regime ($\lambda/d = 10$), as seen in Figure 5b. Furthermore, MoOCl$_2$ provides a broadband achromatic retardation. To quantify this, we use the operational bandwidth as the spectral region where the phase retardance satisfies the ideal quarter-wave condition within a $\lambda/50$ ($\Delta\delta = 7.2°$) tolerance, which is typically sufficient for the industrial quarter-wave plates[9,52]. Under this criteria, 77 nm-thick MoOCl$_2$ delivers achromatic quarter-wave operation spanning the 445 nm to 525 nm and 730 nm to 945 nm spectral ranges (Figure 5c). Therefore, Figure 5 shows that MoOCl$_2$ circumvents the need for complex multi-element compensation schemes[12,14], perovskite heterostructures[9], and costly nanolithography[20] for the creation of highly-efficient polarization elements.

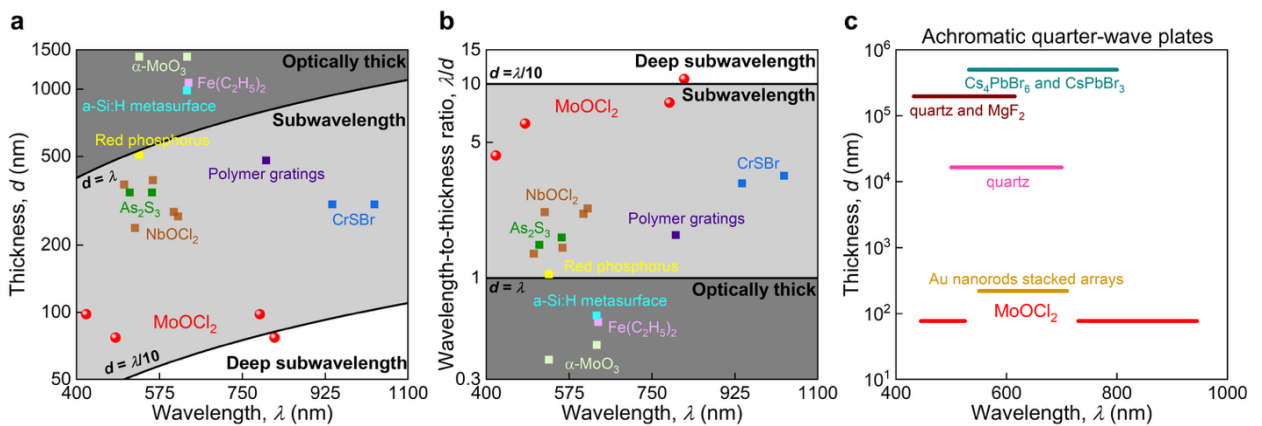

**Figure 5. Benchmarking MoOCl$_2$ quarter-wave plates against state-of-the-art birefringent materials and metasurfaces. a,** Comparison of the required thickness to achieve quarter-wave retardation as a function of the operating wavelength. The solid black lines designate the boundaries for the optically thick ($d = \lambda$) and deep subwavelength ($d = \lambda/10$) regimes. MoOCl$_2$ is plotted alongside other highly anisotropic natural materials, such as CrSBr[34], As$_2$S$_3$[10], $\alpha$-MoO$_3$[33], NbOCl$_2$[35], red phosphorus[36], and Fe(C$_2$H$_5$)$_2$[37] crystals, and artificial nanostructures like polymer gratings[21] and a-Si:H metasurfaces[22]. **b,** The wavelength-to-thickness ratio ($\lambda/d$) across the visible and near-infrared spectrum for the same set of materials. MoOCl$_2$ exhibits exceptionally high $\lambda/d$ ratios, approaching the deep subwavelength regime ($\lambda/d > 10$), demonstrating superior compactness and phase-modulation efficiency compared to both traditional and emerging anisotropic platforms. **c,** Thickness comparison of broadband achromatic quarter-wave plates across their operating spectral ranges. MoOCl$_2$ achieves achromatic phase retardation at deep subwavelength thicknesses, scaling orders of magnitude thinner than traditional bulk crystal combinations (quartz[12] or quartz and MgF$_2$[14]) and competing artificial architectures like stacked Au nanorod arrays[20] and perovskites heterostructures[9].

In summary, we have established MoOCl$_2$ as a highly efficient platform for deep subwavelength and broadband achromatic polarization control. By exploiting MoOCl$_2$ giant optical anisotropy we achieved quarter-wave retardation for the ultrathin 77 nm and 98 nm waveplates. Our findings reveal that the total phase shift in these ultrathin configurations is not merely a product of classical phase accumulation, but is strongly enhanced by Fabry-Pérot cavity resonances within the material. As a result, MoOCl$_2$ waveplates can operate beyond deep subwavelength limit ($\lambda/d > 10$) with up to $\lambda/4500$ retardance tolerance at central wavelengths, outperforming traditional and vdW anisotropic crystals[10,33–37] and engineered metasurfaces[20–22] in both physical footprint and broadband performance.

Despite these exceptional optical characteristics, the transition from fundamental demonstration to practical nanophotonic integration presents several critical challenges and opportunities. Currently, the realization of $MoOCl_2$ waveplates relies on mechanically exfoliated flakes, which inherently restricts the active device area. A pivotal next step will be the development of large-area wafer-scale synthesis techniques for $MoOCl_2$. Drawing inspiration from recent breakthroughs in roll-to-roll mechanical exfoliation for large-area vdW films with preserved crystallographic alignment[53], scalable methods can be optimized for $MoOCl_2$. Furthermore, the dynamic control of $MoOCl_2$ phase retardation remains an exciting frontier. The metallic-dielectric duality of $MoOCl_2$ suggests that its optical anisotropy could be actively tuned. Future investigations should explore the modulation of the dielectric tensor via external stimuli, such as electrical gating[54], thermo-optic gating[55], or strain engineering[56], paving the way for reconfigurable $MoOCl_2$ ultracompact polarization elements.

## Supplementary Information

Supplementary Information contains sections Materials and Methods.

## Author Contributions

G.E. and A.T. contributed equally to this work. G.E., G.T., A.A., K.S.N., and V.V. suggested and directed the project. G.E., A.T., A.S., N.P., A.M., and G.T. performed the measurements and analyzed the data. Z.S. and A.S. synthesized the crystals. V.M. and A.V. provided theoretical support. G.E. and V.M. wrote the original manuscript. All authors reviewed and edited the paper. All authors contributed to the discussions and commented on the paper.

## Competing Interests

The authors declare no competing interests.

## Data Availability

The datasets generated during and/or analyzed during the current study are available from the corresponding author upon reasonable request.

## Acknowledgments

Z.S. was supported by ERC-CZ program (project LL2101) from Ministry of Education Youth and Sports (MEYS) and by the project Advanced Functional Nanorobots (reg. No. CZ.02.1.01/0.0/0.0/15_003/0000444 financed by the EFRR).